\documentclass[5p,times]{elsarticle}

\usepackage{amsmath}
\usepackage{graphicx}
\usepackage{dcolumn}
\usepackage{bm}
\usepackage{hyperref}
\usepackage[mathlines]{lineno}
\usepackage{orcidlink}

\def\gtorder{\mathrel{\raise.3ex\hbox{$>$}\mkern-14mu
 \lower0.6ex\hbox{$\sim$}}}
\def\ltorder{\mathrel{\raise.3ex\hbox{$<$}\mkern-14mu
 \lower0.6ex\hbox{$\sim$}}}
\def\GMn{\ensuremath{G_M^n}}
\def\GEn{\ensuremath{G_E^n}}
\def\GMp{\ensuremath{G_M^p}}
\def\GEp{\ensuremath{G_E^p}}
\def\etal{\textit{et al.}}

\begin{document}


\title{Updated analysis of neutron magnetic form factor and the nucleon transverse densities}

\author[LBNL,JLab]{T.~J.~Hague\orcidlink{https://orcid.org/0000-0003-1288-4045}}
\address[LBNL]{Lawrence Berkeley National Laboratory, Berkeley, California 94720, USA}
\address[JLab]{Thomas Jefferson National Accelerator Facility, Newport News, Virginia, 23606, USA}

\author[LBNL]{J.~Arrington\orcidlink{https://orcid.org/0000-0002-0702-1328}}

\author[UNH]{J.~Jayne}
\address[UNH]{University of New Hampshire, Durham, New Hampshire 03824, USA}

\author[UW]{G.~A.~Miller\orcidlink{http://orcid.org/0000-0003-2443-3639}}
\address[UW]{Department of Physics, University of Washington, Seattle, WA 98195, USA}

\author[UNH]{S.~N.~Santiesteban\orcidlink{https://orcid.org/0000-0001-5920-6546}}

\author[TU]{Z.~Ye\orcidlink{https://orcid.org/0000-0002-1873-2344}}
\address[TU]{Department of Physics, Tsinghua University, Beijing, China}

\date{\today}

\begin{abstract}

We provide an updated global extraction of the neutron magnetic form factor, including new extractions from $^3$H-$^3$He comparisons at Jefferson Lab. Our new global fit addresses discrepancies between previous data sets at modest momentum transfer by separating the uncertainties from world data into normalization and uncorrelated uncertainties. We use this updated global fit, along with previous fits for the other form factors, to extract the neutron and proton transverse charge and magnetization densities and their uncertainties. 

\end{abstract}

\maketitle

\section{Introduction}

Precise measurements of the nucleon electromagnetic form factors over a large range of four-momentum transfer squared, $Q^2$, reveal the spatial distributions of charge and magnetization within the nucleon~\cite{sachs62,arrington07a, perdrisat07, carlson07, arrington11a}. The form factors can be used to extract the model-independent transverse charge and magnetization distributions of the nucleon~\cite{Miller:2007uy, Miller:2010nz,venkat11}. The form factors, evaluated at very low values of $Q^2$ also enter into precise calculations of atomic energy levels that allow tests of the Standard Model~\cite{Pohl:2013yb, Carlson:2015jba, lee15, Borah:2020gte}.

In addition to being connected to the underlying spatial structure of the nucleon, the electromagnetic form factors are also important because they parameterize the elastic e-N scattering cross section. Precise knowledge of the form factors is an important input to many measurements, e.g. in A(e,e'N) single-nucleon knockout measurements, where the e-N cross section has to be accounted for to pull out information on the nucleon distributions in the nucleus.  As such, it is important to have parameterizations of the proton and neutron electromagnetic form factors with realistic estimates of the uncertainties. Ideally these will be well behaved at both low $Q^2$, where they connect to the RMS charge and magnetization radii of the nucleon, and to other low-energy observables, e.g. the hyperfine splitting~\cite{lee15, Borah:2020gte}. It is also important that the parameterization has sensible behavior at high-$Q^2$ values, e.g. for use in future high-energy neutrino and electron scattering measurements.

\begin{figure}[htb]
    \centering
    \includegraphics[width=0.95\linewidth, trim={2mm 3mm 2mm 2mm}, clip]{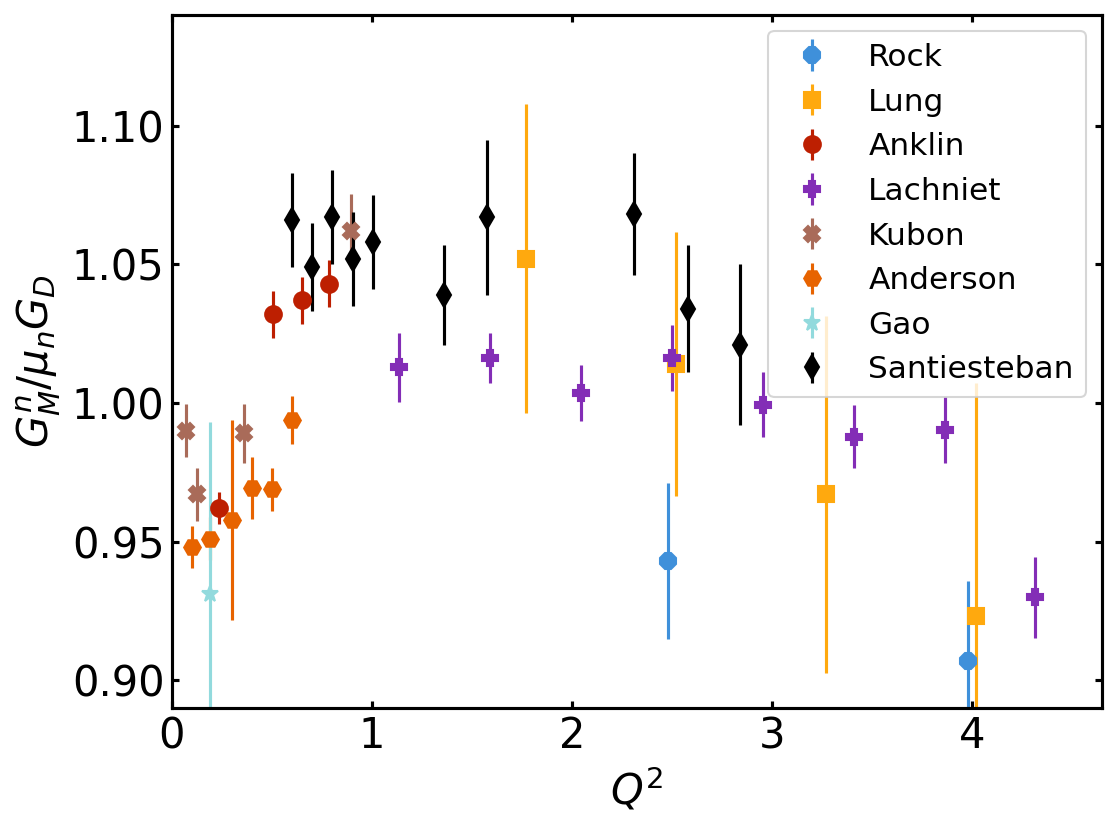}
    \caption{The $\GMn$ data sets included in the previous global fit~\cite{Ye:2017gyb} along with the new results of Ref.~\cite{Santiesteban:2023rsh}. Note that below $Q^2=2$~GeV$^2$, there are multiple high-precision measurements that show significant disagreement.}
    \label{fig:gmn_data}
\end{figure}

A previous global analysis~\cite{Ye:2017gyb} provided parameterizations of the electromagnetic form factors of the proton and neutron, along with their uncertainties, including constraints to provide reliable form factor information for both very small and large $Q^2$ values. For the neutron magnetic form factor, $\GMn$, data sets covering $0.5<Q^2\ltorder$1~GeV$^2$ showed systematic disagreements, as seen in Figure~\ref{fig:gmn_data}. As no clear explanation for the discrepancy was available, the uncertainties of these measurements were artificially enhanced in this region. This was a somewhat arbitrary adjustment that reduced the tension between these data sets but was not based on a detailed analysis of the experiments.  The most recent $\GMn$ extraction from $^3$H-$^3$He comparisons~\cite{Santiesteban:2023rsh} covered a wide $Q^2$ range, including the region where previous measurements yielded inconsistent results. They also demonstrated that the discrepancies could be resolved if a modest normalization uncertainty, taken to be 1.5\%, was added to the previous experiments, most of which did not separate normalization and uncorrelated uncertainties. This observation, along with the new data from~\cite{Santiesteban:2023rsh} which overlaps the $Q^2$ range of multiple data sets, motivated us to update the global fit with this new data and with more realistic estimates of the uncertainties from previous measurements.

In this work, we present an updated fit for $\GMn$ which can be combined with the fit of the other form factors from Ref.~\cite{Ye:2017gyb}. The updated $\GMn$ fit includes an additional data set~\cite{Santiesteban:2023rsh} and an improved breakdown of the systematic uncertainties in previous data sets. We estimate the normalization uncertainty for each of the previous $\GMn$ extractions based on the descriptions of the experiments and analyses of these measurements~\cite{rock82, lung93, gao94, anklin98, kubon02, anderson07, lachniet09}. We also include additional systematic uncertainties in cases where we now have a better understanding of corrections applied in previous measurements, e.g. the e-p scattering contribution in quasielastic measurements on the deuteron, including two-photon exchange (TPE) contributions~\cite{arrington11b}. Using these updated systematic uncertainties, we repeat the global fit procedure of Ref.~\cite{Ye:2017gyb} to provide an updated parameterization of $\GMn$ and its uncertainties. This approach includes several elements that help constrain the behavior at both low and high $Q^2$ values, providing a parameterization with realistic uncertainties. With the updated set of nucleon electromagnetic form factors, we provide an updated extraction of the proton and neutron 2D transverse distributions, following the approach of Ref.~\cite{venkat11} which extracted the proton densities based on an earlier global fit.
on{Form Factor Data Sets}

We start with the same data sets as Ref.~\cite{Ye:2017gyb}, including two high-$Q^2$ constraint points, and add the new $\GMn$ data from Ref.~\cite{Santiesteban:2023rsh}. Following Ref.~\cite{Ye:2017gyb}, we used the updated analysis~\cite{Gao:1998aja} instead of the results from their original publication~\cite{gao94}. We also updated the values of $\GMn$ extracted from Ref.~\cite{rock82} based on the modified analysis of Ref.~\cite{rock92} which updated the $\sigma_n/\sigma_p$ ratio but did not provide an updated extraction of $\GMn$. In addition, we applied a small correction for the contribution of $\GEn$, which was taken to be zero in the original extraction. Finally, the data from Ref.~\cite{lachniet09} are rebinned, as in~\cite{Ye:2017gyb}, because the higher density of points in $Q^2$ will tend to artificially reduce the impact of the systematic uncertainties in the global fit. 

Most of these experiments did not include estimates of the scale uncertainty, treating all experimental and model-dependent uncertainties associated with the extraction of $\GMn$ as uncorrelated or point-to-point (p2p) uncertainties.  Ref.~\cite{Ye:2017gyb} added additional uncertainties for some data sets, including small scale uncertainties, but these were generally minimal and very conservative additions, except in the case of the experiments that showed significant disagreement in the $0.5 < Q^2 < 1.0$~GeV$^2$ range. We perform a more detailed examination of the uncertainties for previous measurements, examining each data set to estimate the appropriate scale uncertainties, as well as any contributions associated with uncertainties neglected in the original extraction, e.g. from TPE~\cite{arrington11b}. Note that, as in the previous analysis~\cite{Ye:2017gyb}, some data sets were excluded ~\cite{markowitz93, anklin94, bruins95} due to concerns about the reliability of the extractions and/or their uncertainties~\cite{jourdan97, bruins97, arrington07a, perdrisat07, carlson07, arrington11a}.

\section{Updated systematic uncertainties}

As noted above, the inclusion of appropriate scale uncertainties is necessary to obtain a reliable fit of the world's data. Ref.~\cite{Santiesteban:2023rsh} showed that a fit to world's data up to $Q^2=3$~GeV$^2$ yielded a good fit and consistency between the different measurements when a 1.5\% scale uncertainty is applied to each of the data sets. This was an initial examination that did not include an explicit evaluation of scale uncertainties. For a reliable global extraction of $\GMn$, it is important to evaluate the uncertainties for each data set individually. We examined the previous $\GMn$ extractions to provide updated estimates of the systematic uncertainties and to separate them into p2p and scale uncertainties. This involved examining multiple potential modifications to the quoted uncertainties:
\begin{itemize}
    \item For uncertainties that were not included in the original analyses, e.g. TPE in e-p and e-n scattering, we estimated and applied additional scale and p2p contributions for each data set.

    \item For systematic uncertainties that are expected to be highly correlated and of similar size across all measurements in the data set, we estimated the fraction that should be treated as scale uncertainty. This contribution is removed from the quoted p2p uncertainties and added as a scale uncertainty. 
    
    \item When a source of uncertainty might be be highly correlated, but is of very different sizes for different $Q^2$ values, adding a large scale uncertainty and removing it from the p2p uncertainties is not appropriate. For these cases, we include a scale uncertainty that is typically below the average uncertainty, but which we estimate to be large enough to allow for reasonable normalization based on the nature of the uncertainty. This uncertainty is applied with no reduction in the quoted p2p uncertainties. This approach allows for the possibility normalization shift while still allowing for significant $Q^2$ dependence, consistent with the originally quoted uncertainties. 
\end{itemize}
Most measurements are directly sensitive to the value of $\sigma_n/\sigma_p$, except for the inclusive measurements which use p(e,e') elastic data to remove the proton contribution from d(e,e') quasielastic measurements~\cite{rock82, lung93}. Therefore, all of the extractions are sensitive to the proton cross section used in extracting $\GMn$, whether it is extracted in the same experiment or taken from global fits. While all extractions except for~\cite{lachniet09, Santiesteban:2023rsh} neglected TPE contributions in the elastic e-p cross section~\cite{arrington11b}, it turns out this generally does not yield additional uncertainty~\cite{arrington04a}. The extraction is sensitive to $\sigma_p$, including TPE contributions, and in nearly all cases, TPE corrections were not applied to the elastic cross sections and so cross section measurements taken during the experiment have the TPE contributions included. The same is true for experiments which used parameterized proton form factors, as long as they were extracted without the inclusion of TPE corrections (and without the inclusion of polarized extractions, where the impact of TPE contributions is different). Earlier measurements neglected TPE contributions altogether and did not have polarization data, while later experiments where TPE corrections were examined~\cite{lachniet09, Santiesteban:2023rsh} specifically chose proton form factor parameterizations that neither remove the TPE contributions nor include recoil polarization measurements. While TPE contributions to the proton cross section do not need to be accounted for, note that experiments that use parameterized e-p cross sections may have additional uncertainties due to any differences (experimental, radiative corrections, etc.) between the data that went into the parameterization and the e-p contribution in their experiments. Such differences can yield errors that are likely to be at least partially correlated across different $Q^2$ values.

Most experiments also neglected TPE contributions to the e-n cross section~\cite{arrington11b}, which yields a small correction in the extraction of $\GMn$. Neutron TPE contributions are generally small and a common uncertainty (0.5\% scale, 0.2\% p2p) was applied to $\GMn$ for all experiments except Ref.~\cite{Santiesteban:2023rsh}, which included a TPE correction and uncertainty in their extraction.

Finally, some experiments have uncertainties that might be significantly correlated between different $Q^2$ values. In many cases, these uncertainties are similar in size for all measurements within a single data set, and we have to estimate what fraction of the uncertainty is correlated. For example, a target thickness uncertainty represents a true normalization uncertainty, so we apply it as a scale uncertainty and remove the full contribution from the p2p uncertainties. For neutron detection efficiency, we expect the uncertainty in the efficiency measurement to be highly correlated, so we assign a significant fraction to the scale uncertainty, removing it from the quoted p2p. 

In other cases, such as the uncertainty associated with modeling nuclear effects, the corrections might be significantly correlated but can also vary substantially with $Q^2$. In these situations, we want to allow for a reasonable scale uncertainty, while still allowing for sufficient variation with $Q^2$. To accomplish this, we take the lower range of the uncertainty and apply this as a scale uncertainty, but do not remove it from the p2p uncertainties, as doing so would not allow points with smaller uncertainties to retain their relative size. While this approach may  be somewhat conservative, it permits a certain amount of correlated error while remaining consistent with the $Q^2$ dependence allowed by the original uncertainties. 

No modifications were made to the results of Santiesteban, \etal~\cite{Santiesteban:2023rsh}, which included estimates of scale and p2p breakdown, for all of the corrections examined.  Modifications to the quoted uncertainties for individual experiments are summarized below and the total added uncertainties are listed in Table~\ref{tab:kinematics}. This summary does not include the common 0.5\% scale and 0.2\% p2p uncertainty associated with neutron TPE contributions. Note that these are quoted as uncertainties for $\GMn$, and where the uncertainties in the original publications are applied to the cross section ratio, we divide the fractional uncertainty by a factor of two to convert them into approximate $\GMn$ uncertainties, as $\sigma_n \propto (\GMn)^2$, neglecting the impact of the small correction from $\GEn$. 

\begin{itemize}
\item Rock,~\etal~\cite{rock82,rock92}: 1\% target thickness uncertainty fully converted to scale; 4\% of model dependence uncertainty converted to scale (from the total 5--20\% model dependence~\cite{rock92}).
\end{itemize}

\begin{itemize}
\item Lung,~\etal~\cite{lung93}: 0.5\% target thickness converted to scale; 2\% scale and 0.6\% p2p added to account for the model dependence, based on the table in Ref~\cite{lung93} showing the impact of using different nuclear models in extracting $\GMn$. 
\end{itemize}

\begin{itemize}
    \item Gao,~\etal~\cite{gao94, Gao:1998aja} The neutron TPE uncertainty was added, but the scale and p2p uncertainties were added in quadrature, since the distinction is irrelevant for the single data point of this measurement. 
\end{itemize}

\begin{itemize}
\item Anklin,~\etal~\cite{anklin98}: Added 0.5\% uncertainty (separated into 0.35\% scale, 0.35\% p2p) to account for the use of older proton form factors in the original analysis. Converted the 0.4\% threshold calibration from p2p to scale.  Converted 0.3\% of the proton loss uncertainty (quoted as 0.3-0.4\%) to scale. Added the following scale uncertainties without reduction of p2p: 0.4\% illumination matching, 0.2\% charge exchange correction, 0.3\% neutron efficiency correction, 0.3\% position-dependence of neutron efficiency, 0.2\% nuclear model, 0.25\% for uncertainty in assumed $\GEn$.
\end{itemize}

\begin{itemize}
\item Kubon,~\etal~\cite{kubon02}: The corrections are identical to those applied to Anklin,~\etal~except that 0.4\% of the proton less uncertainty was converted from p2p to scale, and we added scale uncertainties of 0.4\% for the position-dependent efficiency correction (0.4\%) and 0.3\% for the $\GEn$ model dependence.  
\end{itemize}

\begin{itemize}
\item Anderson,~\etal~\cite{anderson07}: Converted the 0.75\% p2p uncertainty for beam and target polarization to scale. Converted 1\% of the proton form factor~\cite{hohler76} uncertainty from p2p to scale. Converted 1.6\% of the total model-dependence uncertainty (2-5\%) to scale to account for correlated contributions.
\end{itemize}

\begin{itemize}
\item Lachniet,~\etal~\cite{lachniet09}: Converted the following from p2p to scale: 0.7\% of the proton cross section uncertainty, 0.5\% of the losses due to Fermi smearing, and 0.5\% of the $\GEn$ uncertainty. In addition, a 1.5\% scale uncertainty is added for the neutron detection efficiency, but the p2p is only reduced by 0.5\% because there are two detectors covering different angles, yielding two $Q^2$ ranges for each detector at the two beam energies. Thus, a large correlation is possible, but there is also the possibility of significant differences in different $Q^2$ ranges.
\end{itemize}

\begin{table}
\begin{tabular}{|c|c|c|}
\hline
Experiment    & Added uncertainty   & Added uncertainty    \\
              & scale (\% on $\GMn$) & p2p (\% on $\GMn$)    \\ \hline
Rock~\cite{rock82, rock92}  & 4.13     &  -4.12       \\
Lung~\cite{lung93}          & 2.08     &   0.35       \\
Gao~\cite{gao94, Gao:1998aja} & N/A    &   0.54       \\
Anklin~\cite{anklin98}      & 0.98     &  -0.34       \\
Kubon~\cite{kubon02}        & 1.13     &  -0.34       \\
Anderson~\cite{anderson07}  & 2.01     &  -2.00       \\
Lachniet~\cite{lachniet09}  & 1.90     &  -1.05       \\
Santiesteban~\cite{Santiesteban:2023rsh} & - & -  \\
\hline
\end{tabular}
\caption{Summary of systematic modifications for the experiments. These include the common neutron TPE uncertainty and the experiment-dependent modifications listed above.  In most cases, the p2p uncertainty is reduced because uncertainties quoted as p2p have been converted to scale. In Ref.~\cite{lung93}, the addition of a significant model dependence uncertainty led to a net increase in the p2p uncertainty.}  
\label{tab:kinematics}
\end{table}

\section{Fit Procedure}

This global fit uses the $z$ transformation method prescribed in Refs.~\cite{hill10,lee15} as used in Ref.~\cite{Ye:2017gyb}.
This technique is a conformal mapping of $Q^2$ to a unit circle cut on the region of analyticity, that is $t = -Q^2 \leq 4m_{\pi}^2$.
By this construction, the form factor is constrained to a region where it can be fully described by a power series,
\begin{equation}
    G_M^n = \sum_{i=0}^k a_i z^i~,
\end{equation}
where $a_i$ are the fit parameters and $z$ is 
\begin{equation}
    z(t) = \frac{\sqrt{t_{\mathrm{cut}}-t} - \sqrt{t_{\mathrm{cut}}-t_0}}{\sqrt{t_{\mathrm{cut}}-t} + \sqrt{t_{\mathrm{cut}}-t_0}}~,
\end{equation}
where $t_{\mathrm{cut}}$ is the cut value of the analytic region of the form factor and $t_0$ is a free parameter used to set the $Q^2$ mapping of the $z=0$ point.
Cuts on the analytic region are defined by $t_{\mathrm{cut}}=4m_{\pi}^2$, where $m_{\pi}$ is the charged pion mass, and we have chosen to use $t_0=0.7$ GeV$^2$ to match the analysis of Ref.~\cite{Ye:2017gyb}.

In this work, we fit $G_M^n/\mu_n$ directly, where $\mu_n$ is the neutron magnetic moment. We plot our results as $G_M^n/\mu_nG_D$ to more readily highlight the structure of the magnetic form factor, where $G_D$ is the standard dipole form:
\begin{equation}
    G_D = \frac{1}{\left( 1 + \frac{Q^2}{0.71\text{ GeV}^2} \right)^2}.
\end{equation}

\subsection{Fitting Constraints}

We follow a similar prescription to Refs.~\cite{Ye:2017gyb, lee15} in applying constraints to our fit procedure.
These physically motivated constraints require behavior predicted by quark counting rules of the fitted form factor in the $Q^2\rightarrow0$ and $Q^2\rightarrow\infty$ limits.
In the $z$ domain, these limits are
\begin{equation}
    \left.\frac{G_M^n}{\mu_n}\right|_{z\left(Q^2=0\right)}=1
    \label{eq:lowq}
\end{equation}
and
\begin{equation}
    \left.\frac{d^n}{dz^n}G_M^n\right|_{z=1}=0,~~n=0,1,2,3.
    \label{eq:highq}
\end{equation}
These enforce the correct $Q^2=0$ value and proper $Q^2$ asymptotic scaling of $G_M^n\sim Q^{-4}$.
These constraints are applied by defining an $N^{\text{th}}$-order polynomial, but only fitting $N-5$ parameters.
The first parameter is calculated to provide the correct limit at $Q^2=0$, while the last four parameters are calculated to yield the desired behavior at large $Q^2$ (Eq.~\ref{eq:highq}).
These calculated values are included when assessing the residuals of the fit to ensure that the physical constraints do not mar an otherwise good fit to the data.
To further constrain the high-$Q^2$ behavior, we also add the same two pseudo-data points used in Ref.~\cite{Ye:2017gyb} (included in the Supplemental Material~\cite{supplemental}).

We constrain the fitted neutron magnetic radius by adding the PDG value of the neutron magnetic radius ($0.864\pm0.009$ fm~\cite{pdg16}) as an additional data point in the fit. We also apply constraints on the value of each polynomial fit parameter. These take the form of Gaussian priors that $\left|a_k\right|<5$, implemented as in~\cite{Ye:2017gyb}. These bounds are only applied to the polynomial coefficients that are fit; the calculated coefficients are allowed to go to any value that is required to satisfy the constraints.

\section{Results}

We fit the data, using the described techniques, with polynomials from $8^{\text{th}}$ to $17^{\text{th}}$ order. The reduced $\chi^2$ value decreased going from 8 to 10 parameters and was essentially constant for 10 parameters and above. Similarly, the fit result and uncertainties are nearly independent of the number of parameters for $n \ge 10$. As such, we base the fit on the $10^{\text{th}}$ order polynomial, which is also consistent with the fit of Ref.~\cite{Ye:2017gyb}. Our resulting fit has 46 degrees of freedom and a total $\chi^2$ of 58.63 for a $\chi^2$ per degree of freedom of 1.27. 

\begin{figure}[htb]
    \centering
    \includegraphics[width=0.95\linewidth, trim={2mm 3mm 2mm 2mm}, clip]{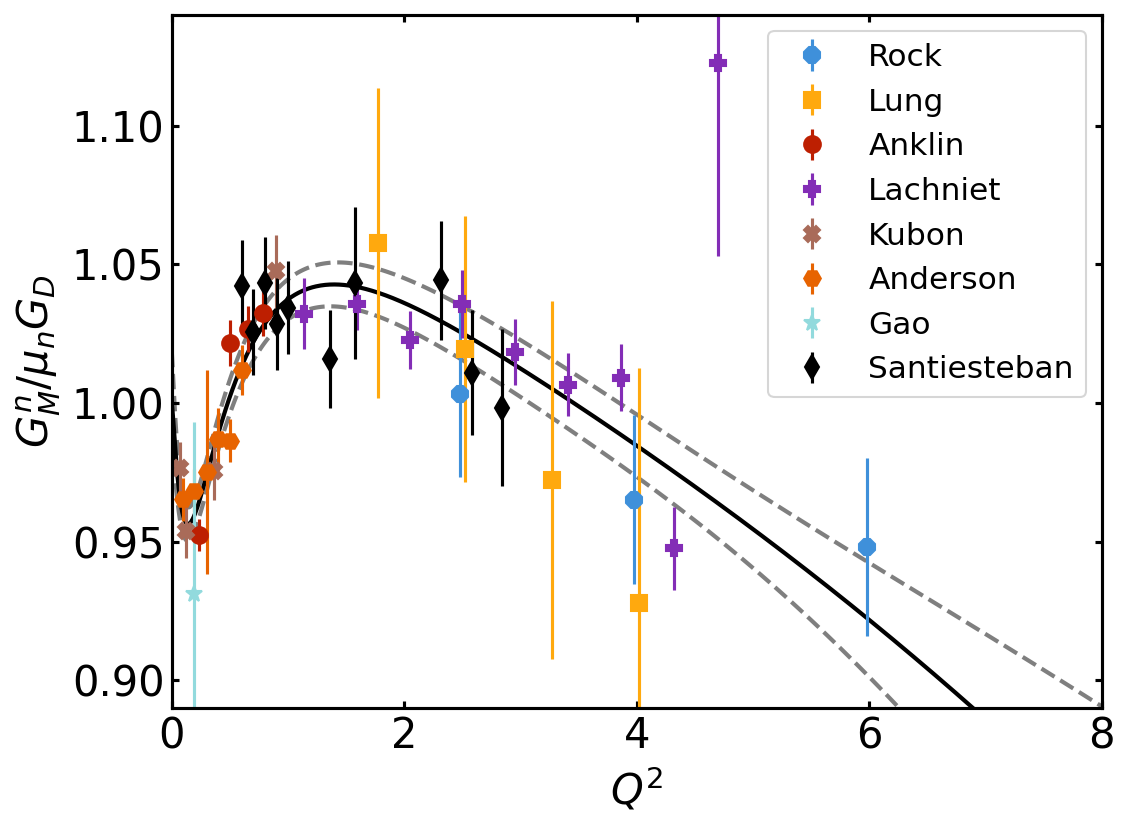}
    \caption{$\GMn$ data included in the fit, after the normalization factors shown in Tab.~\ref{tab:final_norms} have been applied, along with the fit. The solid line indicates the best fit, and the dashed lines indicate the $1\sigma$ uncertainty band}
    \label{fig:fit_linearx}
\end{figure}

\begin{table}[htb]
    \centering
    \begin{tabular}{|l|c|r c l|}
        \hline
        ~Data Set & Scale Unc. & Fit Norm. & $\pm$ & $\delta_{\text{Fit Norm.}}$\\
        \hline
        ~Rock        & 4.1\% & 1.064 & $\pm$ & 0.058 \\
        ~Lung        & 2.1\% & 1.005 & $\pm$ & 0.021 \\
        ~Anklin      & 1.0\% & 0.990 & $\pm$ & 0.017 \\
        ~Lachniet    & 1.9\% & 1.019 & $\pm$ & 0.006 \\
        ~Kubon       & 1.1\% & 0.986 & $\pm$ & 0.010 \\
        ~Anderson    & 2.0\% & 1.018 & $\pm$ & 0.007 \\
        ~Gao         & N/A &  & - & \\
        ~Santiesteban~& 2.4\% & 0.978 & $\pm$ & 0.007 \\
        \hline
    \end{tabular}
    \caption{Fit normalizations and uncertainties. }
    \label{tab:final_norms}
\end{table}

\begin{figure}[htb]
    \centering
    \includegraphics[width=0.95\linewidth, trim={2mm 3mm 2mm 2mm}, clip]{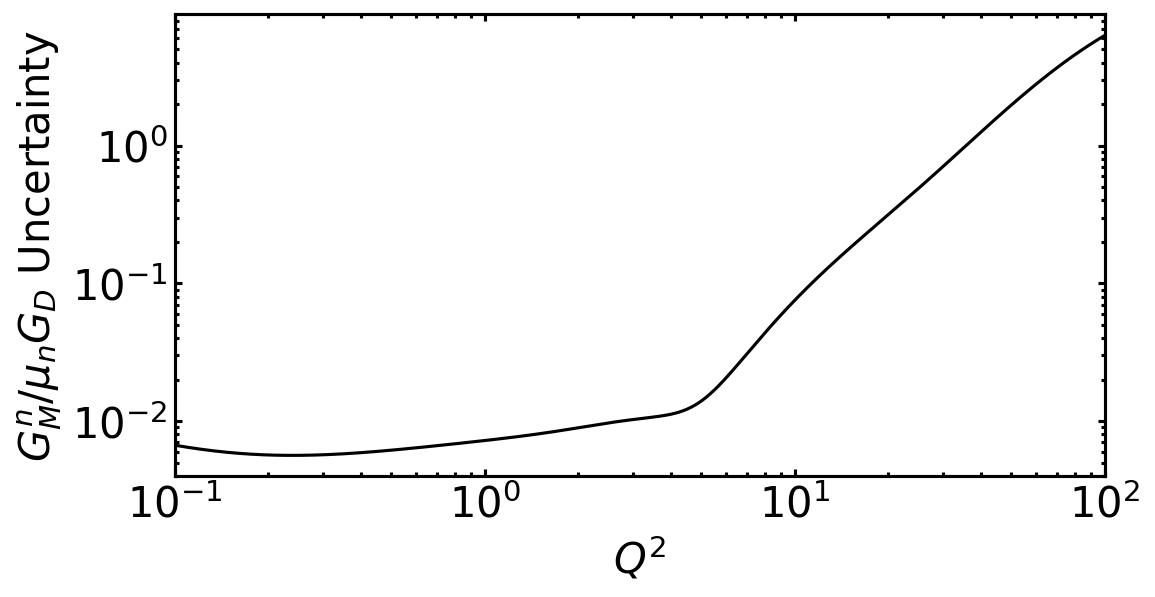}
    \caption{The uncertainty in $\GMn$/$\mu_n G_D$ over a wide range in $Q^2$ values.}
    \label{fig:fit_logx}
\end{figure}

Figure~\ref{fig:fit_linearx} shows the updated $\GMn$ values and the fit from this work. In these plots, the normalization factors from the fit (Table~\ref{tab:final_norms}) have been applied to the data points and so the points are shifted slightly compared to Fig.~\ref{fig:gmn_data}. We note that our fit of $\GMn$ is generally consistent with that from ~\cite{Ye:2017gyb}, with percent-level decreases for $Q^2 \approx 0.1$~GeV$^2$, and few-percent level increases for $Q^2$ from 2-10~GeV$^2$. The uncertainty in the new fit is $\sim$20\% lower over most of the fit region. Figure~\ref{fig:fit_logx} shows the uncertainty from the fit over an extended $Q^2$ range. Code to calculate the fit values and uncertainties using the full covariance matrix are provided in the supplementary materials~\cite{supplemental} along with a simple lookup table.

\section{Transverse densities}

While electromagnetic form factors have historically  been used to probe the underlying charge and magnetization densities of hadrons, traditional three-dimensional Fourier transform methods are not rigorously applicable for systems with constituents that move relativistically, such as the light up and down quarks. The use of the transverse charge density is a rigorously defined way to analyze electromagnetic form factors of hadrons~\cite{Miller:2010nz}. 
The present updated extraction of $\GMn$, combined with the fits for $\GEn$, $\GEp$, and $\GMp$ from Ref.~\cite{Ye:2017gyb}, gives a complete set of fits for the nucleon electromagnetic form factors. These form factors can be used to obtain precise nucleon charge and magnetization densities.

A proper determination of a nucleon charge density requires a relation with the square of a field operator, and this can be achieved through the transverse charge density~\cite{Soper:1976jc, Burkardt:2002hr, Diehl:2002he, Miller:2007uy, Carlson:2007xd}. This is necessary because relativistic effects make extraction of the 3D spatial charge and magnetization densities model dependent~\cite{Miller:2009sg}, while their transverse densities are well defined. 

The charge density $\rho_{ch}(b)$ of a spin 1/2 particle is expressed as a simple integral over the form factors:
\begin{eqnarray}
\rho_{ch}(b)= \int_0^\infty\; {\frac{dQ\;Q}{2\pi}}J_0(Q b) \frac{G_E(Q^2)+\tau G_M(Q^2)}{ 1+\tau},
\label{use}
\end{eqnarray}
with $\tau=Q^2/4M^2$, with $M$ the nucleon mass,  and $J_0$  the  cylindrical Bessel function of order zero. The quantity $\rho_{ch}(b)$ is directly related to the squares of light front wave functions, which are expressed in terms of relative coordinates~\cite{Brodsky:2000ii,Miller:2009sg}.

The magnetization density is given by
\begin{eqnarray}
{\rho}_M(b)= 
\;b \int _0^\infty\frac{Q^2\;dQ}{2\pi}J_1(Qb)\frac{G_M(Q^2)-G_E(Q^2)}{1 +\tau}.
\label{rhotilde}
\end{eqnarray}
In this expression, the angle  between the direction of $\bf b$ and that of the transverse magnetic field, which is also the direction of the nucleon polarization, is taken to be $\pi/2$.  This is in accord with the expectations of classical physics. 
A current in the $z$ direction causes a magnetic dipole density $\sim {\bf r}\times\vec{J}$ in the $x$-direction for positions $\bf r$ along the $y$-direction.  

Ref.~\cite{venkat11} presented a technique to extract the charge and magnetization densities along with realistic estimates of their uncertainties based on parameterizations of the nucleon form factors. This work extracted the proton densities, using a fit of the proton form factors based on Ref.~\cite{arrington07c} but including preliminary results from later high-$Q^2$ polarization transfer extractions of $\GEp/\GMp$~\cite{puckett10}. The proton form factors from Ref.~\cite{Ye:2017gyb} include a full set of high-$Q^2$ polarization measurements, including the updated analysis of recoil polarization extractions~\cite{puckett12, Puckett:2017flj}. This corresponds to a complete analysis of world unpolarized cross section and polarization observables, with the exception of recent unpolarized cross section measurements~\cite{Christy:2021snt}. Inclusion of this data set is complicated by the fact that a different radiative correction procedure was applied, and so a self-consistent analysis requires either the exclusion of this data set or the exclusion of measurements where there is not enough information provided to allow the radiative corrections to be updated, as they were in~\cite{Christy:2021snt}.

Ref.~\cite{venkat11} did not examine the transverse densities of the neutron, and so the densities shown here are the first such results extracted using this technique. We combine the $\GEn$ form factor parameterizations of~\cite{Ye:2017gyb} with the updated $\GMn$ parameterization from this work to provide an extraction of the neutron charge and magnetization densities using an updated set of global fits performed in a consistent framework to parameterize the neutron form factors.

\begin{figure*}[htb]
    \centering
    \includegraphics[width=0.95\columnwidth]{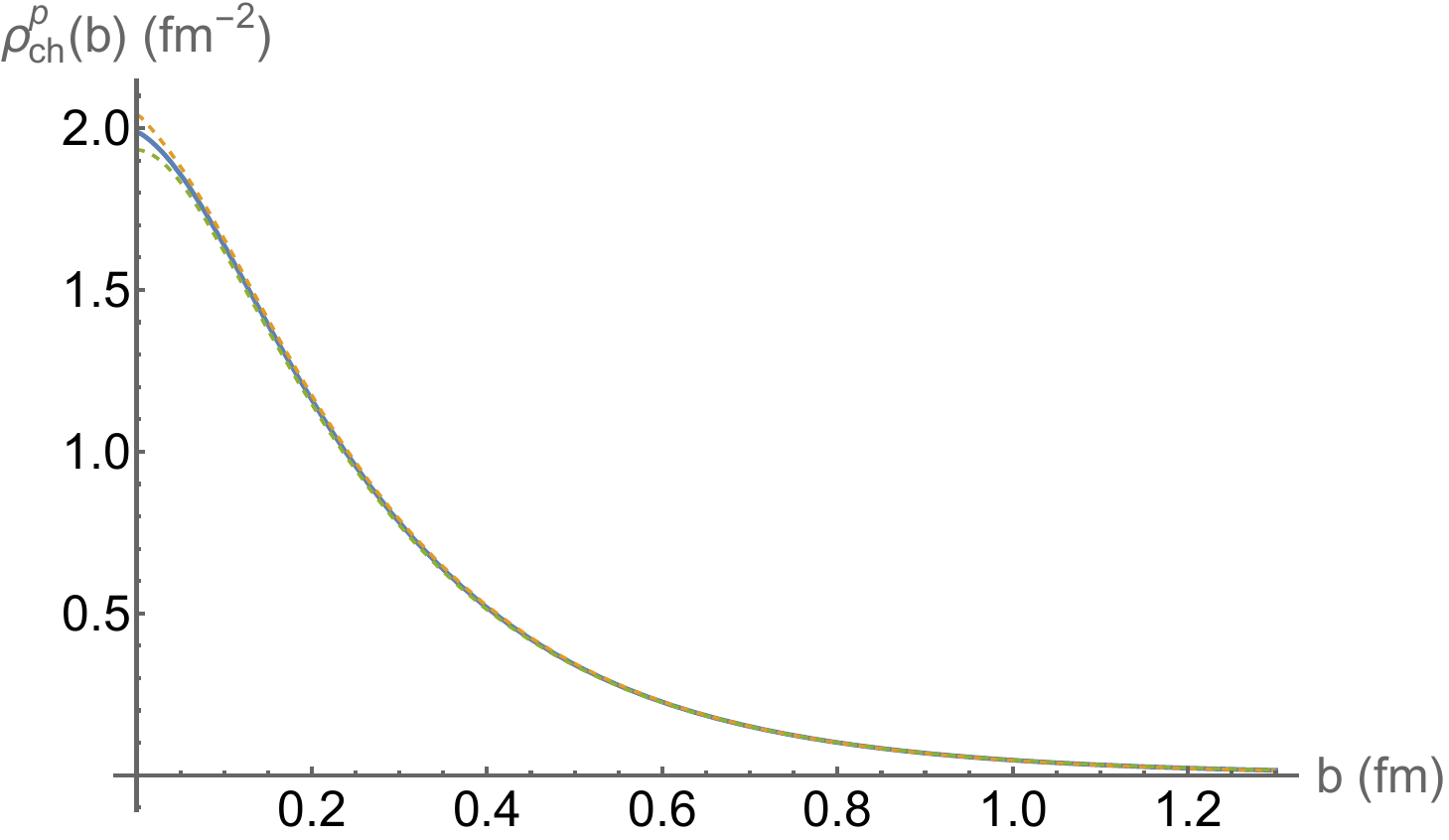} 
    \includegraphics[width=0.95\columnwidth]{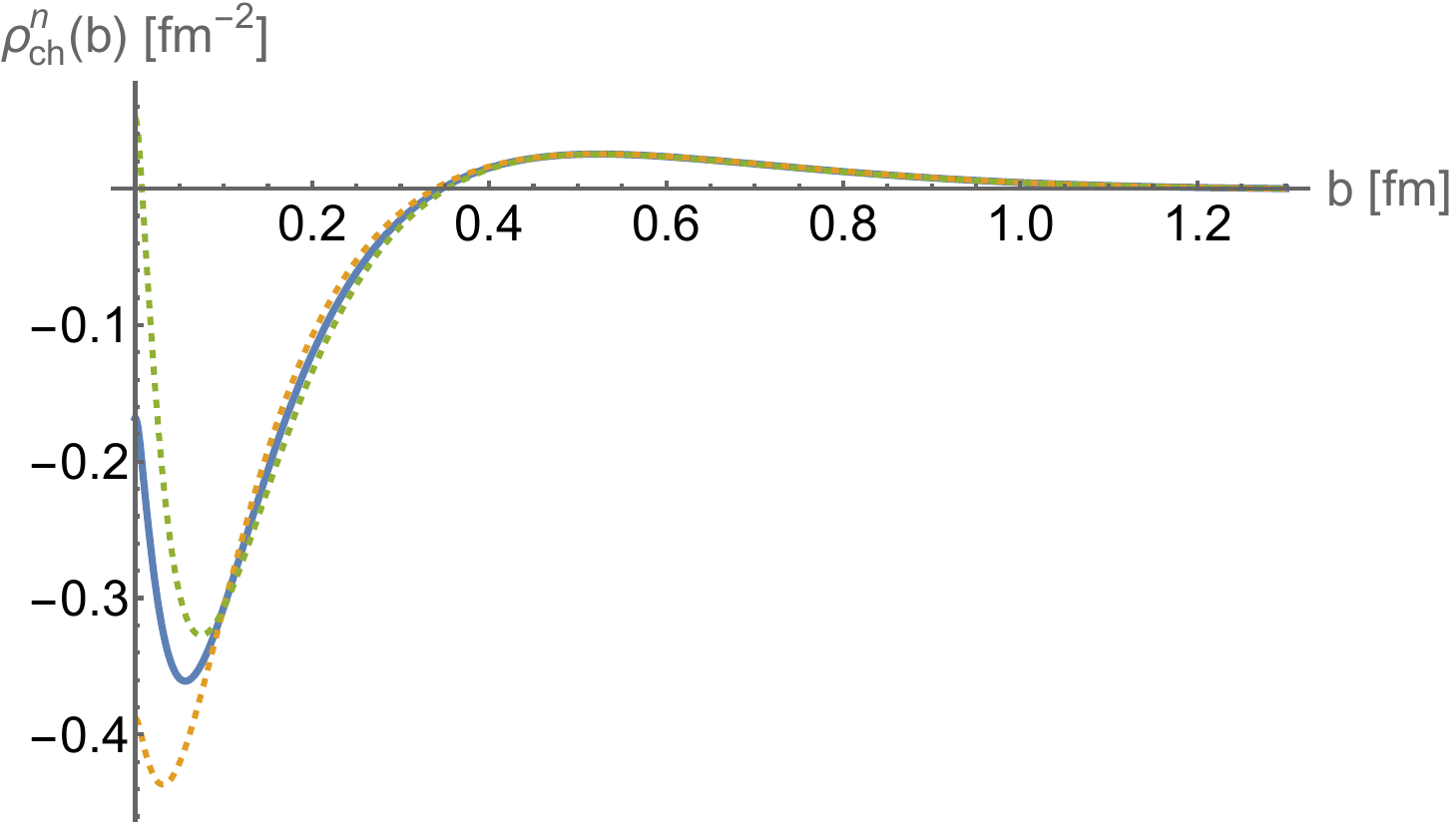} \\
    \includegraphics[width=0.95\columnwidth]{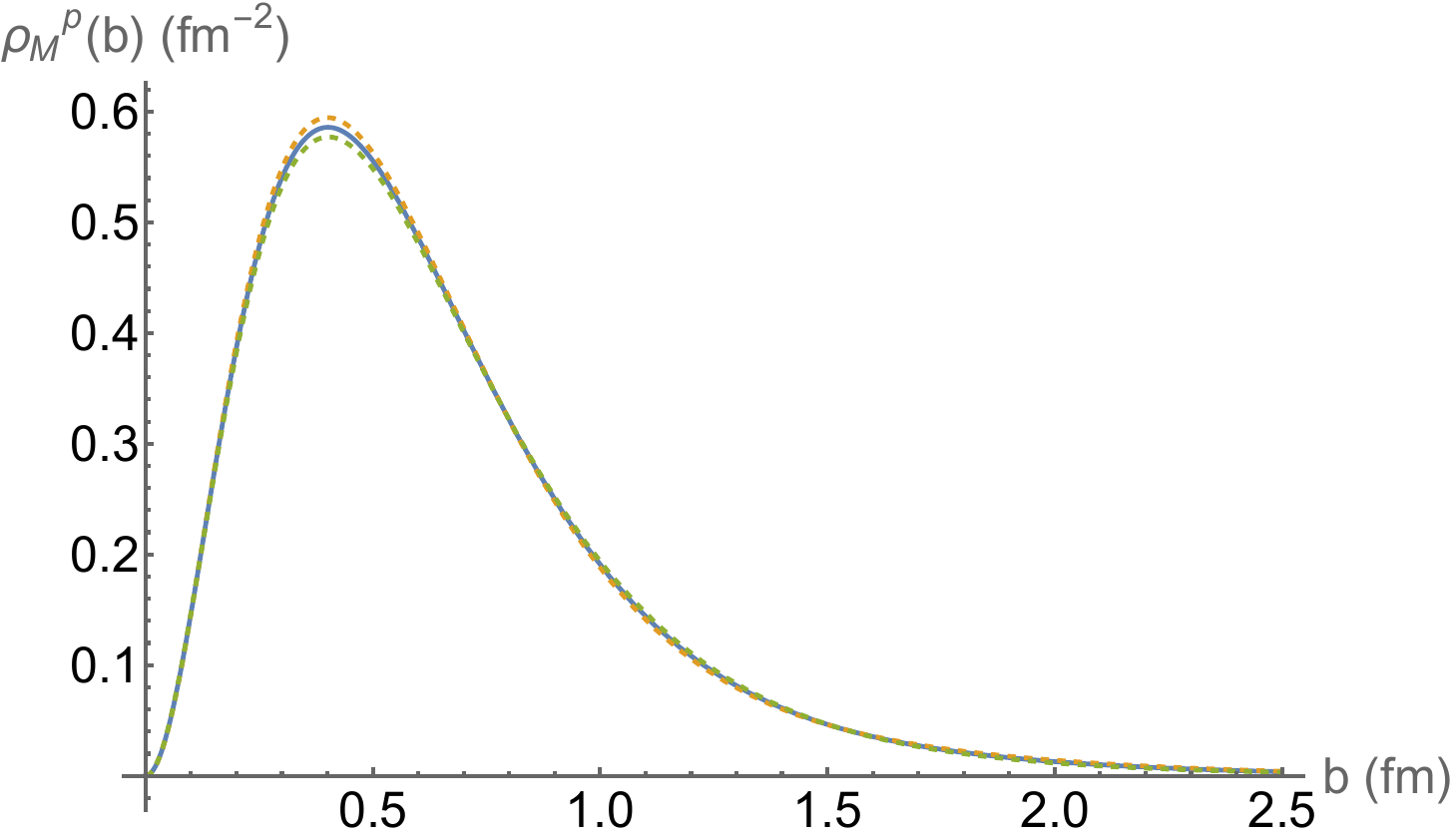}
    \includegraphics[width=0.95\columnwidth]{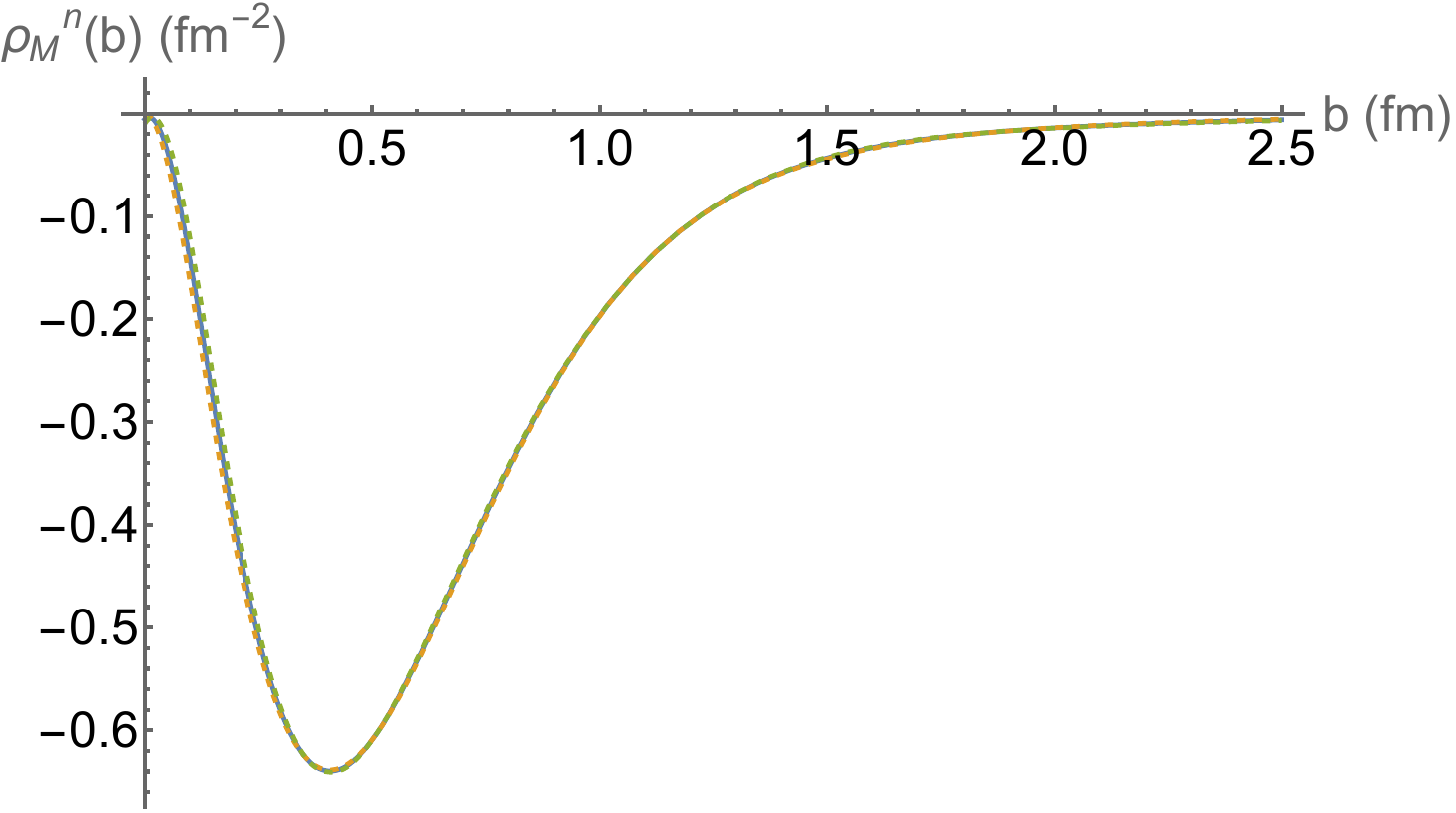} \\
    \caption{The tansverse nucleon charge and magnetic densities; the left (right) panels are for the proton (neutron), while the top (bottom) panels are the charge (magnetic) densities. The solid line indicates the extracted density, while the orange and green dotted lines are the lower/upper error bands. Note that the errors are small making the uncertainties hard to see in most cases. The densities and uncertainties are included in the supplemental material~\cite{supplemental}.}
    \label{fig:densities}
\end{figure*}

Figure~\ref{fig:densities} shows the proton and neutron charge and magnetization densities and their estimated uncertainties. The proton densities are in good agreement with the proton densities extracted in~\cite{venkat11}, with somewhat smaller uncertainties in the charge density. The reduced uncertainty comes from a combination of two factors. The first is theinclusion of additional high-$Q^2$ polarization measurements of $\GEp/\GMp$. In addition, the high-$Q^2$ behavior of these global fits is better constrained than for the fit of Ref.~\cite{arrington07c}, allowing for the fit uncertainties to be used at large $Q^2$ rather than applying a cutoff ($Q^2_{max}=31$~GeV$^2$ as in~\cite{venkat11}.

Concerning the neutron: uncertainties in the charge density at small values of $b$ are significantly larger than for the proton. This is due in part to the limited precision and reduced $Q^2$ range of the neutron charge form factor measurements compared to the proton. In addition, the peak magnitude of the neutron charge density is a factor of 5 below that of the proton charge density, further enhancing the relative uncertainties in the electric and magnetic Sachs form factors. The observed negative  charge density   for small values of $b$   and positive behavior for large values of $b$  is opposite of what has been extracted for the 3-dimensional  charge density~\cite{kelly02}, but is understood to be the impact of the d-quark dominance in the neutron valence region in the transverse infinite-momentum frame densities~\cite{miller08b}. An explanation in terms of a pion cloud model is presented in Ref.~\cite{PhysRevC.80.025206}. 
 
The neutron magnetization density is very similar in form to that of the proton, but is of the opposite sign. The magnitude of the neutron magnetization density is slightly larger than that of the proton because of the larger anomalous magnetic moment. This similarity is expected both from the non-relativistic quark model\cite{Close:1979}, which ignores pion cloud effects, and from models in which the nucleon's pion cloud dominates its structure; for example, Ref.~\cite{Thomas:1981vc} in the limit that bag size is taken to be very small. 

\section{Conclusion}

We examined the uncertainties of previous extractions of $\GMn$, in particular the separation of systematic uncertainties into scale and point-to-point uncertainties, to account for the systematic differences between different experiments. We updated the global fit of Ref.~\cite{Ye:2017gyb}, including both new data on $\GMn$~\cite{Santiesteban:2023rsh} and using the updated uncertainties for earlier measurements. The fit procedure imposes realistic constraints on the form factor at both low and high $Q^2$ values as well as providing realistic estimates of the uncertainties. Allowing for the scale uncertainties in the updated evaluation of the data sets, we obtain a fit to the world data that does not require additional artificial enhancement of the uncertainties, as was done in~\cite{Ye:2017gyb} to provide a good fit. Combining the new parameterization of $\GMn$ with the other form factor results from~\cite{Ye:2017gyb} gives a complete set of fits and uncertainties for the nucleon electromagnetic form factors. 

We then used this set of parameterizations to provide an updated extraction of the transverse charge and magnetization densities for the proton, following the procedure of Ref~\cite{venkat11}, along with a new extraction of the neutron densities. The constraints on the high-$Q^2$ behavior allowed the extraction to be performed without a truncation of the form factor at high $Q^2$ which, combined with the additional data of this analysis and Ref.~\cite{Ye:2017gyb} yields improved uncertainties in the densities compared to the previous extraction. 

\section*{Acknowledgments}
This work was supported in part by the Department of Energy's Office of Science, Office of Nuclear Physics, under contracts DE-AC02-05CH11231, DE-AC05-06OR23177, and DE-SC0024665.
Z. Ye acknowledges the National Natural Science Foundation of China, under grant 12275148, and Tsinghua University Initiative Scientific Research Program. G. A. Miller acknowledges the hospitality of MIT's Laboratory for Nuclear Science.

\bibliography{GMn_bib}
\bibliographystyle{elsarticle-num}

\end{document}